\documentclass[preprintnumbers,twocolumn]{revtex4}
\usepackage{epsfig}

\begin{document}

\begin{abstract}
CeTe$_3$ is a layered compound where an incommensurate Charge
Density Wave (CDW) opens a large gap ($\simeq $~400~meV) in
optimally nested regions of the Fermi Surface (FS), whereas other
sections with poorer nesting remain ungapped. Through
Angle-Resolved Photoemission, we identify bands backfolded
according to the CDW periodicity. They define FS pockets formed by
the intersection of the original FS and its CDW replica. Such
pockets illustrate very directly the role of nesting in the CDW
formation but they could not be detected so far in a CDW system.
We address the reasons for the weak intensity of the folded bands,
by comparing different foldings coexisting in CeTe$_3$.
\end{abstract}

\title{Fermi Surface reconstruction in the CDW state of CeTe$_3$ observed by
photoemission}
\author{V. Brouet$^{1,2,3}$, W.L. Yang$^{1,2}$, X.J. Zhou$^{2}$, Z. Hussain$^{1}$, N.
Ru$^{4}$, K.Y. Shin$^{4}$, I.R. Fisher$^{4}$ and Z.X.
Shen$^{2,4}$}

\affiliation{$^{1}$Advanced Light Source, Lawrence Berkeley
National Laboratory, Berkeley, CA 94720 (USA)\\$^{2}$ Stanford
Synchrotron Radiation Laboratory, Stanford university, Stanford CA
94305 (USA) \\ $^{3}$ Laboratoire de Physique des Solides,
Universit\'e Paris-Sud, B\^at 510, UMR 8502, 91405 Orsay
(France)\\
$^{4
}$ Geballe Laboratory for Advanced Materials and Department of
Applied Physics, Stanford University CA 94305-4045 (USA)}

\date{\today}
\maketitle

Many electronic instabilities, such as charge and spin density
waves, introduce a new periodicity in a solid, which breaks the
lattice periodicity that was used to build the band structure. For
a commensurate instability, it is well known that bands of the
original (``extended'') zone must be folded back into a smaller
(``reduced'') zone according to the new periodicity in order to
establish the equivalency between all reduced zones. Whether this
reasoning still applies to an \textit {incommensurate} periodicity
raises more complex questions \cite{VoitScience00}. Angle resolved
photoemission spectroscopy (ARPES) is the most direct technique to
visualize the dispersion of occupied bands and should then be an
ideal witness for such behaviors. Surprisingly, the new
periodicity in the CDW state is often difficult to detect, as for
example in the model 2D CDW systems of 1$T$-TaS$_2$ \cite
{PilloPRB01} or 2$H$-NbSe$_2$ \cite {StraubPRL99} families. This
is mainly because the intensity of the bands folded according to
the CDW periodicity, called \textit{shadow bands}, which is
proportional to the CDW coupling is often too low to be detected.
The most common ARPES signature of the CDW is then the opening of
a gap, either uniformly on the entire FS or solely on the FS parts
presenting a good nesting. In a few cases, shadow bands have been
observed \emph{but only in the gapped regions}, as in the quasi-1D
systems (TaSe$_4$)$_2$I \cite{VoitScience00} or NbSe$_3$ \cite
{NbSe3SchaferPRL01}.

The fate of the shadow bands in the metallic regions, when there
remains some, is of fundamental interest to understand the impact
of the CDW and its periodicity on the metallic properties. Many
SDW/CDW systems indeed exhibit residual metallicity (as Cr
\cite{FawcettRMP88}, NbSe$_3$ \cite {OngPRB77} or CeTe$_3$ \cite
{Nancy}) because deviation from perfect nesting leaves pockets of
itinerant carriers. The reduction of the size of FS at the
transition has been studied extensively \cite{FawcettRMP88},
especially through quantum oscillations in transport or magnetic
properties that are sensitive to the \textit{area} of certain
sections of the FS pockets (see for example \cite{Organic} for a
discussion of the situation in organic conductors). However, the
location of these FS pockets, as well as their shape, could never
be mapped out directly by ARPES.

We present here an ARPES study of CeTe$_3$, whose goal is
precisely to reconstruct the FS of the CDW state. CeTe$_3$ is made
out of square Te planes alternating with Ce/Te slabs \cite
{RTe3structure}. This yields a pronounced 2D character, where the
electronic properties of the magnetic slab and of the Te planes
are nearly decoupled. Magnetic susceptibility measurements
indicate that the Ce is trivalent \cite{Nancy}. The donated
electrons completely fill the Te $p$ orbitals in the Ce/Te slab
and partially those in the planes \cite{Kikuchi,DiMasiTEMRTe3}.
The metallic Te planes host an incommensurate CDW that was first
detected by transmission electron microscopy (TEM)
\cite{DiMasiTEMRTe3} and that is already present at room
temperature. It is characterized by a very large gap (about
400~meV), an order of magnitude larger than in transition-metal
dichalgogenides, which facilitates the study of certain aspect of
the CDW. Notably, Gweon \textit{et al.} have measured the gap
anisotropy in SmTe$_3$ and shown that it can be well understood in
the framework of a nesting driven CDW \cite{GweonPRL98}. In this
paper, we focus our attention on the modifications of a model
Fermi Surface of one single Te plane, induced by the two following
couplings that tend to set new periodicities.

\noindent i)\textit{\ the transverse coupling between the planes
and the slabs}. The 3D unit cell has a base in the (a,b) plane
rotated by 45$^{\circ }$ and larger by $\sqrt{2}$ compared to the
square unit of a Te plane (Fig. \ref{Folded}d). We will study to
what extent this unit cell is relevant for describing the
electronic properties of one plane.

\noindent ii) \textit{the CDW instability.} The electron-phonon
coupling stabilizes an incommensurate lattice modulation at wave
vector $q_{CDW}\approx 5/7\ast 2\pi /a,$ where
a=4.4~\AA~\cite{DiMasiTEMRTe3,GweonPRL98}.

In both cases, we clearly evidence the folded bands associated to
these couplings and investigate further the topology of the CDW
FS. Indeed, CeTe$_{3}$, like other RTe$_{3}$ compounds
\cite{DiMasiRhoCM94}, remains a fairly good metal despite the
large amplitude CDW, the in-plane resistivity is 50 $\mu \Omega$
cm at 300~K and the residual resistance 1 $\mu \Omega$ cm or less
\cite{Nancy}. We show that, besides the region where the FS is
destroyed by the opening of the large CDW gap, ``pockets'' are
formed between original and folded bands.

\medskip

\begin{figure} [t]
\centerline{
\epsfxsize=0.45\textwidth{\epsfbox{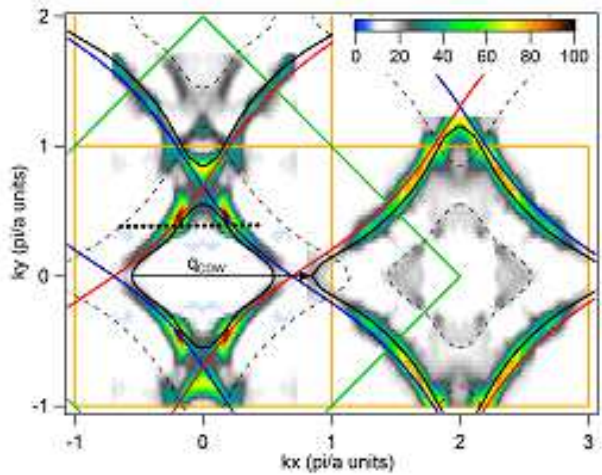}} }
\caption{Map of the spectral weight in CeTe$_3$ obtained at 25 K,
with photon energy h$\nu $~=~55~eV and polarization nearly
perpendicular to sample surface. Orange and green lines indicate
the boundaries
 of the reduced and extended BZ, respectively.} \label{CeTe3FS}
\end{figure}

\begin{figure}[t]
\centerline{
\epsfxsize=0.4\textwidth{\epsfbox{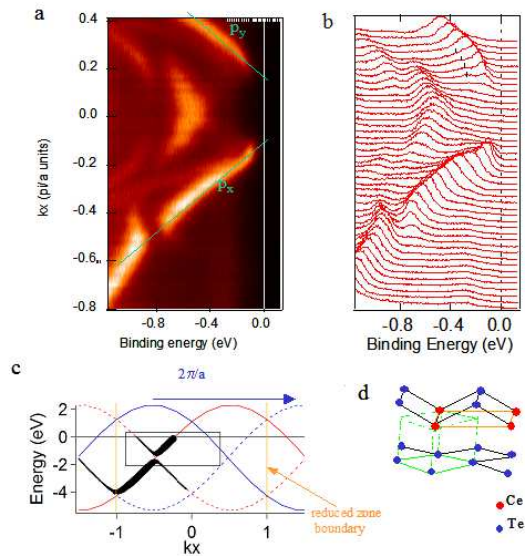}} }
\caption{(a) Color-scale image and (b) EDC stacks of the band
structure for $k_y=0.45~\pi /a.$ (c) Theoretical dispersion for
$p_x$ (solid red line), $p_y$ (solid blue line)
 and the folded bands (dotted lines). The black rectangle correspond to the data shown in a. The thick black
 lines represent the dispersion after letting the main and folded bands interact.
  The thickness is proportional to the ARPES spectral weight. (d) Detail of the stacking between one Te plane and the Ce/Te slab.}
  \label{Folded}
\end{figure}

Large single crystals of CeTe$_3$ were grown by slow cooling a
binary melt \cite {Nancy}. The crystals cleave easily between two
Te planes, providing a good surface quality for ARPES. All the
data were collected at the BL 10.0.1 of the Advanced Light Source
with a Scienta SES-2002 analyser. Fig.~\ref{CeTe3FS} displays a
map of the ARPES spectral weight in
CeTe$_3$ integrated between E$_F$ and E$_F$-200~meV. The main features of this map are very close from those observed in SmTe%
$_3$ \cite{GweonPRL98}, confirming the minor role of the rare
earth in this electronic structure. The solid black lines are
guides to the eye adjusted to the two main pieces in the map : a
``square'' centered at (0,0) and a larger ``outer FS'' part at
(0,2) and (2,0). The lower intensity in the vicinity of the
k$_y$~=~0 axis denotes the presence of the large CDW gap in this
region \cite {GweonPRL98}, which breaks the near equivalency
between $k_x$ and $k_y$. This implies that the long-range ordering
of the CDW is good
enough to create domains at least larger than the beam spot ($\sim $100~$\mu$%
m), as was also the case in previous ARPES and TEM experiments
\cite {GweonPRL98,DiMasiTEMRTe3}.

A very simple electronic structure is expected for a Te plane,
with perpendicular chains of $p_x$ and $p_y$ orbitals playing the
only significant role, since band calculations show that $p_z$ is
completely filled \cite{Kikuchi,Dugdale}. An elementary 2D
tight-binding calculation including only these two orbitals, gives
the following dispersion for $p_x$ ($p_y$ is identical but
perpendicular).

\begin{equation}  \label{disp}
E_{p_x}(\mathbf{k})=t_{\parallel }\cos \left( (k_x-k_y)\frac a2\right)
+t_{\perp }\cos \left( (k_x+k_y)\frac a2\right)
\end{equation}
where the overlap integrals t$_{\parallel }=-4$ $eV$ and t$_{\perp
}=0.75\,eV $ are taken from the calculation of Kikuchi in LaTe$_2$
\cite{Kikuchi}, which contains isostructural Te planes. Fixing
E$_F$ for each orbital to contain 1.25 electrons, we obtain a
contour for the FS shown as red and blue lines on Fig.\ref
{CeTe3FS}, for $p_x$ and $p_y$ respectively. They describe
extremely well the location of the high-intensity
regions in the map, except, of course, at the crossing between $p_x$ and $%
p_y $, where their mutual interaction, totally neglected in our
calculation, separates the square from the outer FS part. This
good agreement leads us to use the dispersion of Eq.~\ref{disp}
below as a guide for backfolding the bands.

Because the system is strongly 2D, it seems natural to obtain a
good description of the electronic structure with a calculation
based on one Te plane. However, the stacking of the planes is such
that the Brillouin Zone (BZ) in the plane for the 3D unit cell
(the orange squares in Fig.~\ref{CeTe3FS}) is different from the
2D BZ of one plane (larger green square), leading to the obvious
problem that the 1st and 2nd BZ of the true 3D structure are not
equivalent for the calculated FS. This equivalency should be
obtained by folding back the bands along the reduced zone
boundaries (or equivalently translating them by 2$\pi /a$, see the
sketch of Fig.~\ref{Folded}c), which gives the dotted black
contours. With data collected over many BZ (in contrast with
\cite{GweonPRL98}), it is clear that the intensity along this
``folded FS'' is drastically reduced, except near $k_y=\pm 1$ or,
to a lesser extent, for the square of the second BZs. This raises
questions about the meaning of the reduced or extended zone scheme
for an experimental technique like ARPES. Voit \textit{et al.
}\cite{VoitScience00} recently clarified this problem by stressing
that \textit{the intensity of a folded
FS in ARPES is proportional to the coupling responsible of the folding}%
. CeTe$_{3}$ clearly illustrates this principle, the lower
intensity of the folded FS is a direct consequence of the weak
coupling between Te planes and the magnetic slab, in other words,
of the 2D character of the system \cite{MEeffects}.

The band structure along the thick dotted line of Fig. 1 is
presented in Fig. 2 to detail the interactions between the main
and folded bands. The two bands dispersing towards the Fermi level
in Fig.~\ref{Folded}a and 2b correspond to $p_x$ and $p_y$ and
form the square; the bands below $E_F$ at k$_x=0$ probably
correspond to the filled $p_z$ orbital and will not be further
discussed. Fig.~\ref{Folded}c indicates the dispersion expected
for all bands after Eq.~1. Although the folded bands are barely
visible in Fig.~\ref{Folded}a and 2b, a break in the dispersion of
the main bands is clearly observed at E= - 0.8 eV, near the
position expected for the crossing between the main and folded
bands. Such a break typically results from the interaction between
the two bands in a perturbation model. The transverse coupling
$V_{3D}(q=2\pi /a)$ opens a gap at the crossing and add a small
admixture of $\left| k+q\right\rangle $ into $\left|
k\right\rangle $ with a weight depending of the coupling strength
and that decreases very fast away from the crossing.
 The new dispersion is shown by black lines in Fig.~\ref{Folded}c with a thickness proportional to this weight, as is
also the ARPES intensity \cite{VoitScience00}. This scheme also
predicts that the intensity of the folded FS will be stronger near
the zone boundary at $k_y=\pm 1$, because the bands cross there
nearer to E$_F$, as is observed.

Let us note that, although there are two Te planes per unit cell,
shifted by a/2 with respect to each other, this stacking does not
define a new unit cell and, therefore, does not induce any
folding. On the other hand, it could split $p_x$ and $p_y $ into
two bands, similarly to the bilayer splitting well-known from
ARPES studies of Bi$_2$Sr$_2$CaCu$_2$O$_{8+\delta }$
\cite{ReviewDamascelli}. This could be the origin of the shoulder
of $p_y$ indicated by black dots in Fig.~\ref{Folded}b. The
intensity of this shoulder strongly depends on photon energy, it
is nearly as intense as the main band at 35 eV.

\medskip

We have discussed this folding in some details, because the
formation of the CDW can be described in identical terms, only the
origin of the coupling is different and the periodicity is
incommensurate. A spontaneous distortion of the lattice at wave
vector $q$ will be stabilized if this allows enough pairs of
states  $\left| k\right\rangle $ and $\left| k+q\right\rangle $ to
lower their energies through the opening of a gap at E$_{F}$. This
is the case here for $q_{CDW}$, which nests the square into the
outer FS part (see Fig. \ref{CeTe3FS}). For clarity, we will refer
in the following to the bands induced by the 3D structure as
``folded bands'' and to those of the CDW as ``shadow bands''. We
can anticipate to observe weak shadow bands at position translated
by $q_{CDW}$ from the main bands that will :

\noindent i) either interact with the main band to create the CDW
gap, if they cross near $E_{F}$ (more precisely, if the distance
between the crossing and $E_{F}$ is smaller than the gap).

\noindent ii) or draw a FS replica that closes the FS pockets,
when the bands cross away from $E_F$, i.e. for poorer nesting.

\begin{figure}[t]
\centerline{
\epsfxsize=0.45\textwidth{\epsfbox{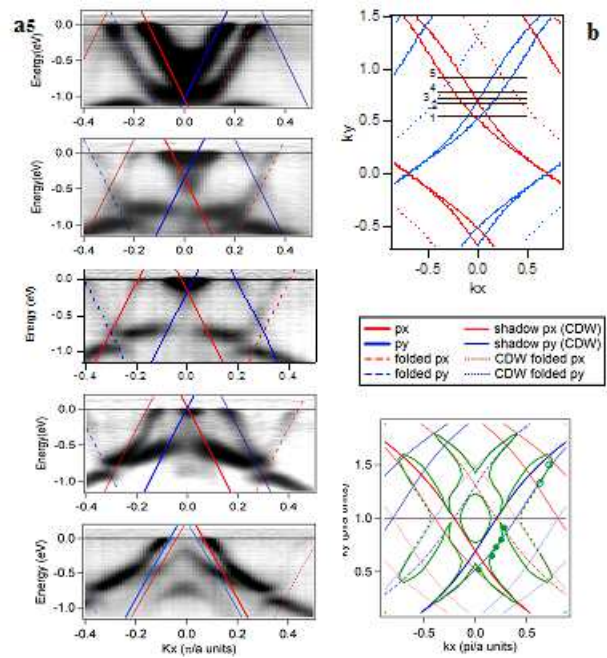}} }
\caption{a1 to a5: Band structure for $k_y$=0.6, 0.69, 0.73, 0.77
and 0.91. A background was substracted and the gray scale is
logarithmic. The lines represent dispersion from Eq.\ref{disp} and
backfolded bands (see legend). b) Location of the cuts of part a
in the reciprocal space with theoretical traces of the main,
folded and shadow FS. c) Sketch of the FS pockets reconstructed
from the different FS traces (green lines, the dashed part is not
observed experimentally). The green points locate the observed
$E_F$ crossings on the shadow FS (solid symbols correspond to
images of part a).} \label{shadow}
\end{figure}

Fig.~\ref{Folded}b gives an example of the first case, that of the
gap formation. It shows that $p_x$ and $p_y$ clearly \textit{bend
away} from $E_F$, leaving a gap $\Delta \simeq 120$~meV (at this
position, the gap has significantly reduced). This bending is due
to the connection with the CDW shadow band, exactly as in the case
previously studied, but with an upper branch, above $E_F$, that
cannot be observed.

The second case, that of FS pockets, occurs in the metallic region
and is
detailed in the images of Fig.~\ref{shadow}. In the bottom image a1, at $%
k_{y}=0.6$, the hole-like square has become ungapped and is nearly
closing. For higher $k_{y}$ values, the electron-like outer FS
piece of FS develops itself around $k_{x}=0$. This evolution
follows very well the theoretical dispersion of Eq.~1, shown as
thick blue and red lines. In this region, the folded outer FS is
also quite clear, giving rise to a second parabola, which becomes
increasingly visible from a1 to a5 (dotted lines). However, two
other weak lines are visible, most clearly in image~a3. They
cannot be attributed to a substructure of another line, such as
the bilayer splitting previously discussed, because their slope is
opposite to that of the next dotted or thick lines \cite{Dugdale}.
\textit{They are the CDW shadow bands}, their position exactly
matches the one of $p_{x}$ or $p_{y}$ translated by $q_{CDW}$
(thin blue or red line) and so does their evolution as a function
of $k_{y}$. The location of the different cuts are indicated in
Fig.~\ref{shadow}b and the Fermi level crossings on the shadow
part of the FS are reported by green points on Fig.~3c. When
approaching the square, the continuity strongly suggests that the
shadow FS smoothly connects to the original square and that the
two bands joining each other just below $E_{F}$ in image a1 (green
dots) actually correspond to the closed CDW replica. However,
bilayer splitting effects become harder to rule out after the
shadow and main bands have crossed and appear parallel.

By connecting the green points, we may start to draw the green
line in Fig.~\ref{shadow}c indicating the contour of the new FS.
To complete this contour, we use the intersection of the main,
folded and shadow FS as a guide. This suggests two different
pockets. The first one, the oval centered at (0,1), is
electron-like and is formed by the interaction between the direct
and folded FS. The second pocket is hole-like and extends from the
top of the square to the outer FS branches. We have discussed its
lower part with Fig.~\ref{shadow}a. Along most part of the outer
FS (shown by dashed green line in Fig.~\ref{shadow}c), we do not
detect shadow bands, although we clearly see a break in the
dispersion where we expect their crossing (just like for the
folded bands in Fig.~2). This lower intensity is expected
theoretically because the nesting is the worst there. At the very
end of the pocket, we observe shadow bands again at positions
indicated by open green circles. This allows to pursue the solid
green line around the closure of the pockets on the outer FS part.

The determination of this green contour was the main purpose of
this paper, as it defines the shape of the metallic pockets.
However, it is only a first approximation of the FS, because we
have restricted our study to the first shadow bands. For the
incommensurate CDW, multiple of $q_{CDW}$ should create an
infinity of new bands. The weight of these higher order shadow
bands decreases very fast, which not only means that they will not
be observable experimentally, but also that their role for many
properties (e.g. transport) will become negligible. Because of
this, the absence of true periodicity introduced by the
incommensurate CDW does not destroy the electronic structure, a
situation comparable to that of quasicrystals \cite
{PiechonPRL95}. Similarly, the spectral weight changes along the
green contour, so that the quasi-classical description of
electronic orbits become non-trivial. However, for each pair of
$\left| k\right\rangle $ and $\left| k+q\right\rangle $ on the
``two sides'' of the pocket, there will be another pair at $\left|
k'=k+q\right\rangle $ and $\left| k'-q\right\rangle $ with
opposite weights, so that the symmetry is maintained. Each
quasiparticles of wave vector k acquires some component at vector
k$\pm $q and interferences between them can occur. Recently, we
have observed quantum oscillations in the magnetization of
LaTe$_3$ \cite{Nancy}, which is a promising step towards a very
complete determination of the RTe$_3$ FS. For fields oriented
perpendicular to the Te planes, we observe osciallations at
approximate frequencies 0.5 and 1.6kT and the former probably
corresponds to the small oval pocket centered at (0,1).

\medskip To summarize, CeTe$_3$ is particularly well suited for a
detailed study of FS topology in presence of an incommensurate
periodicity. The simplicity of the electronic structure of the Te
planes makes it an ideal ground to test the impact of
perturbations. ARPES for example offers a simple image of the way
the 2D electronic structure is modified by the 3D couplings. It
also clearly locates the CDW gap on the best nested parts of the
original FS and further reveals new bands closing the FS that can
be directly traced back to the CDW periodicity. This supports
intimately the description of the CDW as a nesting driven FS
instability, whereas, because of the large gap, it was a priori
not obvious that the CDW could still be viewed as a perturbation
of the metallic state. One could rather have started from a
localized picture, where the tendency of Te to form dimers with
its neighbors would be the driving force of a structural
distortion. Our study clarifies the original metallic properties
arising from this situation, characterized by strong in-plane and
out-of-plane anisotropy. As a result, the FS first appears as made
of \textit {arcs} that are not reaching the zone boundaries. This
situation is reminiscent of that of complex materials exhibiting
pseudogaps in some directions, as in certain phases of cuprates
\cite {ReviewDamascelli}. In CeTe$_3$, we demonstrate, for the
first time in a CDW material, that these arcs can be explained to
a very good approximation as in fact formed of narrow pockets
resulting from the interaction between the original FS and its CDW
replica.

The SSRL's effort is supported by DOE's Office of Basic Energy
Sciences, Division of Materials Science with contract
DE-FG03-01ER45929-A001. The work at Stanford was supported by ONR
grant N00014-98-1-0195-P0007 and  NSF grant DMR-0304981.


\begin{thebibliography}{99}
\bibitem{MEeffects}  With photon energies from 25 to 55 eV and different
polarizations, the relative weight of p$_x$ and p$_y$ changes
significantly but the folded FS is always significantly weaker.
\end{thebibliography}
\end{document}